\documentclass[12pt]{article}
\usepackage{axodraw}
\input epsf.tex
\usepackage[svgnames]{xcolor}

\newcommand{\beq}{\begin{equation}}
\newcommand{\eeq}{\end{equation}}
\newcommand{\beqa}{\begin{eqnarray}}
\newcommand{\eeqa}{\end{eqnarray}}
\newcommand{\beqar}{\begin{eqnarray*}}
\newcommand{\eeqar}{\end{eqnarray*}}

\newcommand{\eps}{\epsilon}
\newcommand{\ga}{\gamma}

\newcommand{\inn}{\!\cdot\!}

\newcommand{\eg}{{\it e.g.,}\ }
\newcommand{\ie}{{\it i.e.,}\ }
\newcommand{\labell}[1]{\label{#1}} 
\newcommand{\reef}[1]{(\ref{#1})}
\newcommand\prt{\partial}
\newcommand\veps{\varepsilon}

\newcommand\cF{{\cal F}}
\newcommand\cR{{\cal R}}
\newcommand\cA{{\cal A}}
\newcommand\cM{{\cal M}}

\newcommand\cJ{{\cal J}}

\newcommand\cL{{\cal L}}
\newcommand\cG{{\cal G}}
\newcommand\cI{{\cal I}}

\newcommand\cP{{\cal P}}

\newcommand\cO{{\cal O}}

\newcommand\bz{\bar{z}}

\newcommand\Tr{{\rm Tr}}

\begin{document}

\vspace*{1cm}

\begin{center}
{\bf \Large
Dilaton Black Hole Entropy \\
from Entropy Function Formalism
}

\vspace*{1cm}

 Komeil Babaei Velni $^\dagger$ \footnote[1]{babaeivelni@guilan.ac.ir},   Ali Jalali $^*$ \footnote[2]{ali.jalali@stumail.um.ac.ir}
  and Bahareh Khoshdelan$^\dagger$ \footnote[3]{bkhoshdelan@gmail.com }\\
\vspace*{1cm}
$^\dagger${Department of Physics, University of Guilan,\\ P.O. Box 41335-1914, Rasht, Iran}
\\
\vspace{0.5cm}
$^{*}${Department of Physics, Ferdowsi University of Mashhad,\\ P.O. Box 1436, Mashhad, Iran}
\\
\vspace{2cm}

\end{center}

\begin{abstract}
\baselineskip=18pt
It has been shown that the entropy function formalism is an efficient way to calculate the entropy of black holes in string theory. We check this formalism for the  extremal charged dilaton black hole. We find the general four-derivative correction on the black hole entropy from the value of the entropy function at its extremum point.    


\end{abstract}
\vskip 0.5 cm


\vfill
\setcounter{page}{0}
\setcounter{footnote}{0}
\newpage
\section{Introduction} \label{intro}
Black hole thermodynamics is a fascinating topic in theoretical physics. From the theoretical point of view, black hole thermodynamics provide an intriguing arena for  quantum gravity researchs. It is expected that a theory of quantum gravity must have an interpretation for the thermodynamic behavior of black holes observed in classical general relativity. So far, several approaches to quantum gravity and semi-classical gravity have discussed this issue\cite{Wald}.
One of them is the Wald’s Noether charge approach that is applicable to general diffeomorphism invariant theories. A quantum field theory calculation of the Wald entropy formula has been presented in \cite{Brown} for general diffeomorphism invariant theories. It shows how this general black hole entropy formula appears from a fundamental
theory of quantum gravity. By explicitly comparing the direct counting of microstates with the Noether charge entropy, the Wald’s approach has been confirmed in many examples in the string theory\cite{Bodendorfer,Brustein}.

It is known that the microscopic description of the black-hole entropy needs the existence of an attractor. The attractor mechanism states that, the radial dependence of the moduli fields given by the equations, whose solutions lead to definite values at the horizon, regardless of their boundary values at infinity. Motivated by this mechanism, Sen has defined the black hole entropy function in higher derivative gravity, in which the Wald formula can be written in terms of this function. Sen's proposal included a particular kind of extremal black holes with the near horizon geometry $AdS_2\times S^{D-2}$\cite{Sen1}. The entropy of such black holes is given by the value of the entropy function at the extremum. Sen has found this function by integrating the Lagrangian density over the horizon and then carry out the Legendre transform of the result with respect to some parameters. 

In \cite{Sen2},\cite{Ghodsi} and \cite{entfunc}, the thermodynamics of some extremal black holes have been investigated by this mechanism. It has been shown that the entropy function formalism works correctly both in ten and lower dimensions. The higher derivative corrections to entropy and also corrections to background at near horizon has been studied. Several related works are given in \cite{Alishahiha}. The higher derivative terms may  modify the solution such that the near horizon is not $AdS_2\times S^{D-2}$. In such cases, the entropy function formalism is not applicable\cite{Garousi1}.

In this paper we would like to show that in the context of $N=4$ supergravity  and for  extremal dyonic dilaton black holes, the entropy function formalism works. We apply this method to find the higher derivative correction to entropy coming from 
general four-derivative terms including the metric curvature and the gauge field strength, added to usual supergravity action.

 An outline of the paper is as follows. In section 2, we review the dyonic dilaton black holes  as a special solution of a dimensionally reduced superstring theory.  In sections 3, using the entropy function formalism,  we compute the entropy of the extremal  doyonic dilaton black holes and show that this formalism works. We apply this formalism in section 4 to find the higher derivative correction to entropy and explicitly only check  the Gauss-Bonnet contribution of these correction terms.

\section{Dilaton black holes}

Understanding  the nature of singularities in gravitational theory and the quantum thermal properties of processes near black holes may be possible with the help of investigation of evaporation of black holes. Dilaton black holes could be the stable endpoints of the evaporation process.

It is shown that $N=4$ supergravity can be consequence of a dimensionally reduced superstring theory in $d =4$. A solution of this reduction can be the spherically symmetric electrically and magnetically charged dilaton
black hole that includes the classical Schwarzschild and Reissner-Nordstrom black holes and dilaton black holes with either purely electric, purely magnetic, or both charges considering in the solution\cite{Kallosh1}. The bosonic part of the action that can be described the above theory is given by
\beqa
I = \int d^4x\,\sqrt{-g}\left( -R +2\partial^\mu\phi\cdot\partial_\mu \phi-e^{-2\phi} {\cal F}_{\mu\nu} {\cal F}^{\mu\nu} \right)\labell{I}
\eeqa
where
\beqa
\cF_{\mu\nu}&=&F_{\mu\nu}+G_{\mu\nu},\nonumber\\
F_{\mu\nu}&=&\partial_{\mu} A_{\nu} - \partial_{\nu} A_{\mu},\nonumber\\
G_{\mu\nu}&=&\partial_{\mu} B_{\nu} - \partial_{\nu} B_{\mu}.\nonumber 
\eeqa
 The dilaton $\phi$ is the real part of a complex scalar in which it's imaginary part is the axion putting to constant. The equations of motion and their solutions have been discussed in \cite{Kallosh1,Gibbons1}. The field strengths $F_{\mu\nu}$ and $G_{\mu\nu}$ satisfy the axion field equation provided that one considers each vector field $A_{\mu}$ and $B_{\mu}$ has to be either electric or magnetic.

The dilaton black holes with electric or magnetic charges have been investigated in \cite{Garfinkle}. In the electric and magnetic dilaton black hole solution, $A_{\mu}$ and $B_{\mu}$ are purely electric and magnetic, respectively. Asymptotically non-vanishing dilaton field $\phi_0$ and the dilaton charge $ \Sigma$ are defined by the equation $\phi \sim  \phi_0 +\Sigma/r$ at $r\rightarrow \infty$. The dilaton charge can be calculated in terms of black hole mass, electric and magnetic charges as $\Sigma =(P^2 - Q^2)/{2M}$ in which $ Q = e^{-\phi_0}Q_{elec}$ and $P = e^{-\phi_0}P_{magn}$.

Consider the extremal solution of the equation of motion of the actin \reef{I} that can be given in the following form\cite{Kallosh1}
\beqa
ds^2=  \frac{(r - M)^2}{r^2-\Sigma^2} dt^2 - \frac{r^2-\Sigma^2}{(r - M)^2} dr^2 -(r^2-\Sigma^2)  d\Omega_2^2.\labell{ds2}
\eeqa
This solution has a duality symmetry that exchanges simultaneously the electric charge and magnetic charge, the dilaton charge and negative sign of dilaton charge. For the above extremal dilaton black hole, there is the following condition between the independent parameters as following
\beqa
 M ^2 + \Sigma ^2= P^2+ Q^2\labell{cs}
\eeqa
It is shown in \cite{Kallosh1} that the bound deriving from supersymmetry is exactly the lower bound on the dilaton black hole mass imposed by cosmic censorship \reef{cs}. It could be found some other solutions by considering the both fields $F$ and $G$  have electric and magnetic charges. In this set of solutions, there are two electric and two magnetic parameters corresponding to the fields $F$ and $G$. For a constant axion and considering the axion field equation, this set of solutions depend on five parameters.
We are interested in the solution of the purely electric extremal dilaton black hole. 
 

\section{Entropy of dilaton black hole}

As it is well known, the entropy function formalism is an applicable method for deriving the entropy of black holes with near horizon geometry $AdS_2\times S^{D-2}$.
Let us review this method here:
consider an extremal black hole with  the near horizon geometry $AdS_2\times S^{D-2}$ in the space-time dimension D in which the $AdS_2$ part is proportional to $-r^2dt^2+dr^2/{r^2}$. The background of the black hole includes different scalar, electric and magnetic fields $u_s,e_i$ and $p_a$ respectively. These fields accompanying with $v_1$ and $v_2$ denoting the sizes of $AdS_2$ and $S^{D-2}$, characterize the background. Define an entropy function by integrating the Lagrangian density over the horizon $S^{D-2}$ and then carry out the Legendre transform of the result with respect to $e_i$. Extremizing this function with respect to the scalar and sizes parameters, the values of these parameters can be determined. Eventually, the value of the result function at the horizon will be corresponded to the entropy\cite{Sen1}.

 In this section, we are going to calculate the entropy of extremal dilaton black hole with two electric charges using the entropy function formalism. The near horizon solution of metric equation of motion \reef{ds2} has the geometry $AdS_2\times S^2$, so the entropy function formalism must be applicable. To apply this formalism, we write the near horizon solution \reef{ds2} in terms of parameters $v_1$ and $v_2$ as following
\beqa
ds^2= v_1\bigg(- \frac{r ^2}{M^2-\Sigma^2} dt^2 + \frac{M^2-\Sigma^2}{r^2} dr^2\bigg) +v_2(M^2-\Sigma^2)  d\Omega_{2}^{2},\nonumber\\
 F = \frac{Q\, e^{\phi_0}}{(M - \Sigma)^2}\  dt \wedge dr,  \,\,\,\,\,\,\,\,\,\,\,\,\,\,\,\,G=P \, e^{\phi_{0}}\, sin \theta \, d\theta \wedge d\phi, \,\,\,\,\,\,\,\,\,\,\,\,\,\,\,e^{2 \phi} =e^{2 \phi_0}\  \frac{M+ \Sigma}{M  - \Sigma}.
\eeqa
One can assume the following values for form fields and the constant value of the scalars near the horizon
\beqa
F_{rt}=e_1,\,\,\,\,\,\,\,\,\,\,\,\,\,\,\, \tilde{G}_{rt}=e_2,\,\,\,\,\,\,\,\,\,\,\,\,\,\,e^{-2\phi}=S.
\eeqa
where we use the duality rotation\footnote{This implies that $\tilde{B_{\mu}}$ is also electric, and the calculations are often simpler when using the electric solution $\tilde{B_{\mu}}$, rather than the magnetic $B_{\mu}$.}
 \beqa
\tilde{G}^{\mu\nu}=\frac{1}{2}i(-g)^{-\frac{1}{2}}e^{-2\phi}\epsilon^{\mu\nu\alpha\beta}G_{\alpha\beta}. \nonumber
 \eeqa

For this background, the ricci scalar becomes $R=2\big(v_2-v_1\big)\big/\big(v_1v_2(M^2-\Sigma^2)\big)$ and the non-vanishing components of the Riemann tensor are
\beqa
&&R_{abcd}=\frac{1}{v_1(M^2-\Sigma^2)}(g_{ac}g_{bd}-g_{ad}g_{bc})\,,\,\,\,\,a,b,c,d=r,t,\cr &&\cr
&&R_{ijkl}=\frac{-1}{v_2(M^2-\Sigma^2)}(g_{ik}g_{jl}-g_{il}g_{jk})\,,\,\,\,\,i,j,k,l=\theta,\phi.
\eeqa
 The vanishing of the covariant derivative of the fields (gauge field strengths, Rieman tensor and scalar fields)  in the near horizon geometry is a main property of the general form of the background using in the entropy function formalism.
 
 The integral of Lagrangian density over the horizon that we denote by $f$ becomes
 \beqa
 f&=&\frac{1}{16\pi}\int d\theta d\phi \sqrt{-g}\cL\nonumber\\
 &=&\frac{v_1-v_2}{2}+\frac{v_2}{v_1}\frac{M^2-\Sigma^2}{2}\bigg(Se_1^2+S^{-1}e_2^2\bigg).
 \eeqa
Extremizing the above function with respect to $S$ and $v$ gives the dilaton and metric equations of motion at the limit of near horizon, respectively. The electric charges are given by the gauge field equations of motion $q_i={\partial f}/{\partial {e_i}}$. By the  Riemann  tensor rescaling $R_{rtrt}\rightarrow \lambda R_{rtrt}$, one can find the entropy as
\beqa
S_{BH}=2\pi (M^2-\Sigma^2)\frac{\partial f_{\lambda}}{\partial {\lambda}}\bigg|_{\lambda=1}\labell{sbhf}
\eeqa 
where the $f_{\lambda}$ is a function similar to $f$ except that the $R_{rtrt}$ rescaling has been done. The variation of Lagrangian density with respect to  Riemann tensors gives the following constraint equation
\beqa
\int d\theta d\phi \sqrt{-g}R_{abcd}\frac{\partial L}{\partial R_{abcd}}=2 \frac{\partial f}{\partial\lambda}\bigg|_{\lambda=1}.
\eeqa
It could be easily found that the above expression is exactly equal to $f-e_1{\partial{f}}/{\partial e_1}-e_2{\partial{f}}/{\partial e_2}$. The entropy function $F$,   that is defined as the Legendre transform of $f$ with respect to electric fields, then becomes
\beqa
F&=&e_i\frac{\partial{f}}{\partial e_i}-f =-\frac{\partial f}{\partial\lambda}\bigg|_{\lambda=1}\nonumber\\
&=&\frac{v_2-v_1}{2}+\frac{v_1}{2v_2}\frac{1}{M^2-\Sigma^2}\bigg(S^{-1}q_1^2+Sq_2^2\bigg).
\eeqa
where we have used the  gauge field equations of motion. By solving  the equation of motion for metric and scalar, one can fix the parameters as following
\beqa
v_1=v_2=1,\,\,\,\,\,\,\,\, S=\frac{q_1}{q_2}.
\eeqa
Substituting the above values and considering the entropy \reef{sbhf}, the entropy in terms of entropy function becomes
\beqa
S_{BH}&=&-2\pi (M^2-\Sigma^2) F=\pi (M^2-\Sigma^2).
\eeqa
It is straightforward to find the above entropy by direct computing the area of the horizon. So we have shown that the entropy function formalism works here. 

\section{Higher derivative correction}
It is well known that one can not use simply the area-entropy law to find the entropy of black holes when the higher derivative correction terms have been taken into account. 
 It is common to organize the interactions by their dimension or alternatively by the number of derivatives\cite{Myers}. The Lagrangian \reef{I} contains covariant terms up to two derivatives.  So it is natural to next consider the possible interactions at fourth order in derivatives.  One can find the following general four-derivative Lagrangian:
\beqa
\Delta \cL=\frac{S}{16\pi G}&\bigg[&\alpha_1 R^2+\alpha_2 R F^2+\alpha_3 R  {\tilde{G}}^2+\alpha_4 (F^2)^2+\alpha_5 F^4+\alpha_6 (\tilde{G}^2)^2+\alpha_7 (\tilde{G})^4\nonumber\\
&&+\alpha_8 R_{\mu\nu}R^{\mu\nu}+\alpha_9 R^{\mu\nu}F_{\mu\alpha}{F^{\alpha}}^{}_{\nu}+\alpha_{10} R^{\mu\nu}{\tilde{G}}_{\mu\alpha}{{{\tilde{G}{}}^{\alpha}}^{}}_{\nu} \nonumber\\
&&+\alpha_{11} R_{\mu\nu\alpha\beta}R^{\mu\nu\alpha\beta}+\alpha_{12} R_{\mu\nu\alpha\beta} F^{\mu\nu}F^{\alpha\beta}+\alpha_{13} R_{\mu\nu\alpha\beta} {\tilde{G}}^{\mu\nu}{\tilde{G}}^{\alpha\beta}\bigg]\labell{cor}
\eeqa

where $F^2=F_{\mu\nu}F^{\mu\nu}$, $F^4={F_{\mu}}^{\nu}{F_{\nu}}^{\alpha}{F_{\alpha}}^{\beta} {F_{\beta}}^{\mu}$ (similarly for $\tilde{G}^2$ and $\tilde{G}^4$) and the $\alpha_i$ are some unspecified coupling constants\footnote{In a string theory context, it might be expected all of these interaction terms to emerge in the low-energy effective Lagrangian as string-loop or $\alpha'$ corrections to the two-derivative supergravity Lagrangian\cite{Myers}. In such a context, these terms would appear as a perturbative expansion where the higher order terms is suppressed by powers of, e.g., the ratio of the string scale over the curvature scale. It has been demonstrated in \cite{Myers} that within a perturbative framework, one can use field redefinitions to reduce the most general four derivative action to include lower interaction terms than appearing in \reef{cor}.}. This induces the change in the functions $f$ and $F$ in which the relation between $q_i$ and $e_i$ changes. This is in contrast to the situation in \cite{Sen2} where the function $f$ corresponding to the correction term is independent of the parameters $v_1$ and $v_2$ and hence the entropy contribution from the correction terms can be calculated separately. So in this case, one should found first the entropy function $F$ corresponding to the lagrangian $\cL+\delta\cL$ and after the corresponding entropy. 

Finally, the contribution of the correction terms to the entropy could be easily determined. The higher derivative terms respect the symmetry of the solution of the original theory, \i.e., the coefficients $v_1$ and $v_2$ remain constant. After finding the equation of motion of the four-derivative Lagrangian \reef{cor} accompanying the original Lagrangian \reef{I}, we see that the extremal black hole with near horizon geometry $AdS_2 \times S^2$  saturate the equation of motion and provided that :
\beqa\labell{Id}
&\alpha_{1}=\alpha_{3},\,\,\,\,\ \alpha_{8}=-4\alpha_{11},\,\,\,\,\ \alpha_{5}=-2\alpha_{4},\,\,\,\,\ \alpha_{7}=-2\alpha_{6}\nonumber\\
&\alpha_{9}=2\alpha_{12}+2\alpha_{2},\,\,\,\,\,\,\,\,\,\ \alpha_{10}=2\alpha_{13}+2\alpha_{3}.
\eeqa 

So one can apply the entropy function formalism for the black hole solutions obtained from the higher derivative lagrangian \reef{cor}.

 The contribution of the higher derivative terms to the function $f$ becomes
\beqa
\Delta f&=&-\frac{C_3S}{2(M^2-\Sigma ^2)}(\frac{v_2}{v_1}+\frac{v_1}{v_2})-\frac{(M^2-\Sigma ^2)}{4}S\frac{v_2}{v_1^3}(C_1e_1^4+C_2e_2^4)-\frac{S}{v_1}(\alpha_2e_1^2+\alpha_3e_2^2)\nonumber\\
&&-\frac{v_2}{2v_1^2}S (C_4e_1^2+C_5e_2^2)+2\frac{\alpha_1}{M^2-\Sigma ^2}S.
\eeqa
The electric charges carried by the black hole are given by
\beqa
q_1&=&\frac{\partial (f+\Delta f)}{\partial e_1}=-Se_1\bigg(\frac{v_2}{v_1^2}C_4+2\frac{\alpha_2}{v_1}\bigg)-(M^2-\Sigma ^2)Sv_2\bigg(\frac{e_1^3}{v_1^{3}}C_1-\frac{e_1}{v_1}\bigg)\nonumber\\
q_2&=&\frac{\partial (f+\Delta f)}{\partial e_2}=-Se_2\bigg(\frac{v_2}{v_1^2}C_5+2\frac{\alpha_3}{v_1}\bigg)-(M^2-\Sigma ^2)\bigg(\frac{Se_2^3}{v_1^{3}}C_2-\frac{e_2}{v_1S}\bigg).\labell{q}
\eeqa
where we define $C_i$'s, the combinations of coupling constants, as following:
\beqa
&&C_1=4\alpha_4+2\alpha_5,\,\,\,\,\,\,\, C_2=4\alpha_6+2\alpha_7,\,\,\,\,\,\,\,C_3=2\alpha_1+\alpha_8+2\alpha_{11},\nonumber\\
&& C_4=-2\alpha_2+\alpha_9+2 \alpha_{12},\,\,\,\,\,\,\,C_5=-2\alpha_3+\alpha_{10}+2\alpha_{13}.\labell{comb}
\eeqa

Following the entropy function formalism, one can calculate the entropy function $\cF$ as the legendre transform of $f+\Delta f$ with respect to the electric fields $e_1$ and $e_2$.
\beqa
\cF&=&(M^2-\Sigma ^2)\bigg(\frac{v_2}{2v_1}(Se_1^2+S^{-1}e_2^2)- \frac{3v_2}{4v_1^3}S(C_1e_1^4+C_2e_2^4)\bigg)\nonumber\\
&&+(M^2-\Sigma ^2)^{-1}S\bigg((\frac{v_2}{2v_1}+\frac{v_1}{2v_2})C_3-2\alpha_1\bigg)-\frac{1}{2}(v_1-v_2)\nonumber\\
&&-\frac{S}{v_1}\bigg(\alpha_2e_1^2+\alpha_3e_2^2+\frac{v_2}{2v_1}(C_4e_1^2+C_5e_2^2)\bigg).
\eeqa
By variation of  $\cF$ with respect to $v_1$, $v_2$ and $S$ and also using \reef{q}, the equation of motion could be found. With the assistance of a package for Mathematica, 
 as well as a symbolic computer algebra system for such problems,  we have found the  solutions of equations of motion  as following:
\beqa
S&=&\frac{3}{2}(M^2-\Sigma ^2)^2\bigg(C_1^3+C_2^3+\frac{1}{2}C_1C_2\bigg)(q_1^2+q_2^2)\nonumber\\
&&+4(M^2-\Sigma ^2)^{-2}\bigg(q_1^2C_3^2+q_1\frac{(M^2-\Sigma ^2)^2}{4}+q_2^2C_3^{\frac{3}{2}}\bigg)\bigg(q_1^{\frac{1}{2}}C_3+\alpha_1^{\frac{1}{2}}q_2^2+4(C_3^{2}+1)\bigg)^{-\frac{1}{2}}\nonumber\\
&&+\frac{1}{24}\bigg(q_1^3\alpha_2^4+q_2^3\alpha_3^4+(C_4^2+C_5^2+4C_4C_5)(q_1^{\frac{1}{2}}-q_2^{\frac{1}{2}})\bigg),\nonumber\\
v_2&=&(M^2-\Sigma ^2)^{-1}\bigg([q_2^2+4(\alpha_2\alpha_3+1)^2]^{\frac{1}{2}}+[\frac{q_1^2(q_2^4+q_1q_2C_3)}{q_2^2(1+C_3^2)+4}]^{\frac{1}{2}}\bigg)\bigg(1-(M^2-\Sigma ^2)C_3\frac{q_1^{\frac{3}{2}}q_2^{\frac{1}{2}}}{1+C_3}\bigg)^{\frac{1}{2}},\nonumber\\
v_1&=&\bigg(q_1^{\frac{1}{2}}q_2^{\frac{3}{2}}C_4C_5[(v_2^2-1)^{-1}-1]-2v_2(\alpha_2(q_1^3+2q_1)+\alpha_3(q_2^3+2q_2))\bigg)^{\frac{1}{2}}\nonumber\\
&&\times\bigg(2q_1^3q_2^3(C_4^2+C_5^2)(v_2^2-1)^{-1}+(v_2^2-1)(C_3+C_3^2)+v_2^{-2}\bigg)^{-\frac{1}{2}}.\labell{sv1v2}
\eeqa

We treat these coupling constants as dimensionless coefficients that have real finit values.  By considering these features of the coupling constants, one can find that solving the above system of equations of motion  produces the following identities for the coupling constants combinations $C_i$'s.
\beqa
&&q_1^{\frac{3}{2}}C_1^2+q_2^{\frac{3}{2}}C_2^2+2q_1^{\frac{5}{2}}C_4^{\frac{1}{2}}+2q_2^{\frac{5}{2}} C_5^{\frac{1}{2}}=0,\nonumber\\
&&(q_1^{-\frac{1}{2}}+q_2^{\frac{1}{2}})(C_1^4+C_2^4+\frac{2}{3}C_1^2C_2^2)+(q_1^{\frac{1}{2}}+q_2^{-\frac{1}{2}})(C_4^2+C_5^2+\frac{4}{3}C_4C_5)=0.\labell{Idd}\nonumber\\
\eeqa
Inserting the values in \reef{sv1v2} into entropy function $\cF$, one can find that, for the interaction we have considered, the entropy takes the form  
\beqa
S_{BH}&=&\frac{\pi}{12}(M^2-\Sigma ^2)^{-2}\bigg(C_3(4C_3+q_1^{\frac{1}{2}}+q_2^2)+1\bigg)^{-\frac{1}{2}}\nonumber\\
 &&\times \bigg(1-\frac{(M^2-\Sigma ^2)C_3q_1^{\frac{3}{2}}q_2^{\frac{1}{2}}}{1+C_3}\bigg)^{\frac{1}{2}}\bigg( 4(1+\alpha_2\alpha_3)^{2}+q_2^2
+\frac{q_1^2C_3(q_1q_2+q_2^2)}{4+(1+c_3^2)q_2^2}\bigg)^{\frac{1}{2}}\nonumber\\
&&\times\bigg[18(M^2-\Sigma ^2)^4(2\alpha_1^{\frac{1}{2}}+2C_1^3+C_1C_2+2C_2^3) C_3(q_1^2+q_2^2)\nonumber\\
&&\times\bigg(C_3(4C_3+q_1^{\frac{1}{2}}+q_2^2)+1\bigg)^{\frac{1}{2}}+96(C_1^3+C_3^{\frac{3}{2}}q_1^2q_2^2)\nonumber\\
&&+(M^2-\Sigma ^2)^2\bigg(C_3(4C_3+q_1^{\frac{1}{2}}+q_2^2)+1\bigg)^{\frac{1}{2}}\bigg(24q_1+2(C_4^2+4C_4C_5+C_5^2)\nonumber\\
&&\times (q_1^{\frac{1}{2}}-q_2^{\frac{1}{2}})+\alpha_3^4q_2^3]\bigg)
\bigg].\labell{Entropy}
\eeqa 

It should be noted that we have not use the identities \reef{Idd} to find the above entropy. When these identities take in to account, the entopy appears with a smaller parameter content.

The Gauss-Bonnet correction is the most popular candidate to imitate string corrections to the Einstein action. This correction can be deduced by only considering the first terms in the three lines of \reef{cor} in which $\alpha_1=\alpha_{11}=1$ and $\alpha_{8}=-4$  \footnote{Setting $\alpha_{8}=-4$ as the only non-zero term, yields the heterotic correction that deduced from the heterotic-type $I$ duality.}.  We calculate the Gauss-Bonnet corrected entropy independently and find it as following:
\beqa
S^{(GB)}_{BH}=2\pi q_1q_2 \bigg(1+4q_2^{-2}\bigg)^{\frac{1}{2}}.
\eeqa
From \reef{comb}, one can easily find that the $C_i$'s would be zero for the Gauss-Bonnet correction. It could be found that the equation of motion corresponding to the Gauss-Bonnet correction and the above corresponding entropy are exactly equal to \reef{sv1v2} and \reef{Entropy}, respectively, when $\alpha_2=\alpha_{3}=0$, $\alpha_{1}=1$ and the all coupling combinations $C_i$ were set to zero. 

We have found the entropy of extremal electrically charged dilaton black hole and calculated the correction entropy in the presence of a general four derivative correction Lagrangian from the entropy function formalism. It would be interesting to find these results
by counting the number of degeneracy of the microstates. It is also interesting to do the same calculation for the non-extremal charge dilaton black hole solution and find  the correction to the tree-level solution by using the entropy function formalism.\footnote{In applying the non-extremal result to extremal case, one must define the entropy of an extremal black hole to be the limit of the entropy of the associated non-extremal black hole in which the non-extremal parameter goes to zero\cite{Mehran}.}


{\bf Acknowledgments}:  We would like to thank A.Ghodsi, M.R Mohammadi Mozaffar, D.Mahdavian Yekta and M.H.Vahidinia for very valuable discussions. This work is supported by University of Guilan.

\end{document}